\documentstyle[pra,eqsecnum,aps,epsfig,amssymb]{revtex}
\begin{document}
\draft
\title{Helium in superstrong magnetic fields}
\author{O.-A. Al-Hujaj and P. Schmelcher}
\date{December 21, 2002}
\address{Theoretische Chemie, Institut f\"ur Physikalische Chemie der Universit\"at Heidelberg, INF 229, 69120 Heidelberg, Germany}
\maketitle
\begin{abstract}
   We investigate the helium atom embedded in a superstrong magnetic field
   $\gamma=100$-$10000$~au. 
  All effects due to the finite nuclear mass for vanishing  pseudomomentum
  are taken into account. 
  The influence and the magnitude of the  different finite mass
  effects are analyzed and discussed. 
  Within our full configuration interaction approach
  calculations are performed for the magnetic
  quantum numbers M=$0$,$-1$,$-2$,$-3$, singlet and triplet states, as well as
   positive and negative z~parities. Up to six 
  excited states for each symmetry are studied. With increasing field
  strength the number of bound states decreases rapidly and we remain
  with a comparatively small number of bound states for
  $\gamma=10^4$~au  within the symmetries investigated here.
\end{abstract}
%\narrowtext

\section{Introduction}

The term ``strong'' field characterizes a situation for which the
system is in the 
nonperturbative regime, i.e. where the magnetic
forces are of the same order of magnitude or greater than the
Coulomb binding force. For the ground state of the hydrogen atom this
corresponds 
to field strengths $\gamma\gtrsim1$~a.u. (1 a.u. corresponds to
$2.35\times 10^5$~T). We refer to the term ``superstrong'' to
indicate a  field strength of $100$ atomic units and more.

The motivation to study atoms and molecules in strong magnetic fields
originates from several sources. Certainly
the properties of atoms and molecules in strong magnetic fields are
interesting from a pure theoretical point of view. Due to the 
competition of the spherically symmetric Coulomb potential and the
cylindrically symmetric magnetic field interaction we encounter a
nonseparable, nonintegrable problem already for a one-electron system,
i.e. the hydrogen atom. 
Therefore it is necessary to develop new
techniques to solve the Schr\"{o}dinger equation in strong magnetic fields.

 The discovery of strong magnetic
fields on the surface of magnetic white dwarfs ($10^2$--$10^4$~T) and
neutron stars ($10^4$--$10^7$~T) is a further major motivation.  The
spectra  of these 
astrophysical objects can be dominated by the 
influence of magnetic and electric fields.
For the analysis of
atmospheres of magnetic white dwarfs and neutron stars, it is very important to
have reliable data  on the behavior of matter in strong magnetic fields.
As an example we mention the  the
white dwarf GrW+70$^\circ$$8247$.
The interpretation of its spectrum was very important for the
understanding of the properties of spectra of magnetic white dwarfs in general
(see Refs. \cite{Angel:1985_1,Angel:1978_1,Greenstein:1985_1,Wunner:1985_1,Wickramasinghe:1988_1}). 

Highly accurate data are available for hydrogen in strong magnetic
fields (see e.g.
\cite{Ruder:1994_1,Kravchenko:1996_1}). This 
system is now understood to a very high degree. But there is also
a significant interest in 
 accurate data on heavier elements such as He, Na, Fe and
molecules. Especially helium plays an important role in the atmosphere
of certain magnetic white dwarfs (see e.g.
\cite{Jordan:1998_1,Jordan:2001_1,Wickramasinghe:2000_1}).

There were several attempts to calculate accurate energies for bound
states of helium, but  for astrophysical applications an
accuracy of approximately $10^{-4}$ for the energies is needed for a large number of
levels. We will concentrate here on 
investigations that address the high field regime:
In 1975 Mueller et al. \cite{Mueller:1975_1} calculated the few lowest
levels of He for $\gamma$ 
up to 20000 a.u. using  a variational approach. Virtamo \cite{Virtamo:1976_1}
presented Hartree-Fock 
calculations on the ground state (which is a triplet state with magnetic
quantum number equal to $-1$). The same state has been considered by
Pr\"oschl \cite{Proeschl:1982_1} in 1982 in the range  21~--~21000~au. Vincke and Baye \cite{Vincke:1989_1} provide 
correlated calculations ($\gamma$=$4.2$, $42$ and $420$ a.u.) for the
lowest singlet and triplet states with positive z parity and magnetic
quantum numbers $M$=$0$,$-1$,$-2$. In the work of Thurner
\cite{Thurner:1993_1} several triplet states are considered in the very
broad range
$\gamma{}$=$8\times10^{-4}$~---~$8\times10^{3}$~a.u. 
We mention also the important work by Becken and Schmelcher,
which covers the same symmetries and the same number of excited states
in the broad range $\gamma=8\times10^{-4}$~---~$100$~a.u. of the
magnetic field strength published in a 
series of papers \cite{Becken:1999_1,Becken:2000_1,Becken:2001_1,Becken:2002_1}.

The properties of matter in superstrong magnetic fields are especially
interesting for the physics of cold neutron stars
\cite{Pavlov:1995_2,Pavlov:1995_1,Taylor:1993_1}. But for the case of
superstrong magnetic fields the  finite nuclear mass effects become
increasingly important. This is due to the fact, that energy shifts,
caused by the finite
nuclear mass,  are of the order of $\gamma/M_0$, where $M_0$ is the
mass of the nucleus. Evidently this correction is
for superstrong magnetic fields of the same order of magnitude as
the ionization energies.

Of course the conceptual and in particular the computational situation
becomes more complex when the full Hamiltonian, i.e. for finite
nuclear mass, is taken 
into account. For both neutral \cite{Schmelcher:1994_1,Johnson:1983_1}
as well as charged systems
\cite{Schmelcher:1994_1,Johnson:1983_1,Baye:1982_1,Baye:1986_1,Vincke:1988_1,Vincke:1990_1,Baye:1990_1,Schmelcher:1991_1,Schmelcher:1995_1,Bezchastnov:1998_1}
we  encounter couplings among the different electrons as well as
couplings between the collective and electronic motion.

Some approximations to the ionic problem are available: An approximate separation of
the collective  and relative motion of the charged system has been
introduced in \cite{Baye:1982_1} and applied to calculate the finite
nuclear mass corrections for low-lying levels of hydrogenic ions in a
magnetic field \cite{Baye:1986_1}. Elaborated MCHF computations as well as
adiabatic approximations were given in Ref. \cite{Bezchastnov:1998_1}.
However these approaches cannot be used to calculate accurate results for
all field strengths and all states investigated here. We mention that
the behavior of the ion He$^+$ has to be known in order to determine
the ionization energies of the neutral He atom.

In the present paper we provide a full configuration interaction
calculation for helium in 
superstrong magnetic fields. All finite mass effects at zero
pseudomomentum are taken into account and are analyzed. In section
\ref{sec:problem} we will 
describe the Hamiltonian, and  some technical details
concerning our calculation. Furthermore we provide some remarks on the
problem of the threshold energies. In section \ref{sec:finite_mass} we
analyze the deviations of the Hamiltonian in the infinite nuclear
mass frame from the full Hamiltonian. Ionization energies and transition
wavelengths, resulting from our calculations, are provided in
Sect. \ref{sec:results}.

\section{Formulation of the problem}
\label{sec:problem}

\subsection{Hamiltonian and Symmetries}

To investigate the He atom we use a nonrelativistic approach. This is well
justified by the fact, that relativistic corrections have been shown to be
very small in  strong and even superstrong magnetic fields
\cite{Chen:1992_1,Poszwas:2001_1}. For 
the sake of simplicity we take the electronic  
spin g factor to be 2, but our results can be easily adapted to any g factor
by multiplying the spin operators and their eigenvalues by $g/2$. On
the other hand, the ionization energies, and transition wavelengths,
which are presented in this paper are not affected by this choice. The
magnetic field vector will be denoted by $\bbox{B}$,  whereas  its
magnitude  will be denoted by
$\gamma=|\bbox{B}|$. The magnetic field vector $\bbox{B}$ is chosen to point
in z~direction. 

The first step in our approach is the pseudoseparation of the collective
and relative motion for the Hamiltonian in the laboratory frame
\cite{Schmelcher:1994_1,Johnson:1983_1,Avron:1978_1} which exploits
the conservation of the 
so-called pseudomomentum $\bbox{K}$. 
The resulting transformed Hamiltonian is divided into three
parts, which are denoted by $H_1$, $H_2$, and $H_3$. The operator
$H_1=\bbox{K}^2/(2M_A)$ involves only center of mass (CM) degrees of
freedom, where $M_A$ is the mass of the atom. $H_3$ contains exclusively 
electronic degrees of freedom. The  operator $H_2=\bbox{K}/M_A\cdot
\bbox{B}\times\sum_i \bbox{r}_i$ 
represents the coupling between $H_1$ and $H_3$, i.e. the CM and
electronic degrees of freedom. 
It involves the motional electric field $1/M_A(\bbox{B}\times\bbox{K})$
which arises due to the motion of the (neutral) atom in the magnetic
field and is oriented perpendicular to the magnetic field.
This coupling is proportional to  the pseudomomentum, and therefore
vanishes
for vanishing pseudomomentum. The pseudoseparation is possible for
neutral systems only, since only then all components of the
pseudomomentum commute.

Within the present work we  assume a vanishing pseudomomentum. 
Therefore we have no additional motional electric field and all
effects due to the finite nuclear mass are included in $H_3$ which is
 a
function of the mass of the nucleus $M_0$ and the magnetic field
$\gamma$. 
In atomic units the electronic Hamiltonian takes the
following form (internal coordinates are taken with respect to the nucleus):
\begin{equation}
 H(M_0,\gamma) = H_{rm}+H_{mp}\\
 \label{eqn:hamiltonian}
 \end{equation}
where
\begin{eqnarray}
 H_{rm}&=& \label{eqn:ham_red_mass}
 \sum_{i=1}^2\left(\frac{1}{2\mu}\bbox{p}_i^2+\frac{1}{2\mu'}\bbox{B}\cdot\bbox{l}_i+\frac{1}{8\mu}(\bbox{B}\times\bbox{r}_i)^2-\frac{2}{|\bbox{r}_i|}+\bbox{B}\cdot\bbox{s}_i\right)+\\
 & &+\frac{1}{|\bbox{r}_1-\bbox{r}_2|}\label{eqn:elec-elec}
\end{eqnarray}
and 
\begin{equation}
  \label{eqn:mass_pol}
H_{mp}=\frac{1}{2 M_0}\sum_{i\not =j}\left(\bbox{p}_i\cdot \bbox{p}_j-\bbox{p}_i \cdot(\bbox{B}\times\bbox{r}_j)+\frac{(\bbox{B}\times\bbox{r}_i)\cdot(\bbox{B}\times\bbox{r}_j)}{4}\right).
\end{equation}
The reduced masses are $\mu=1/(1+1/M_0)$ and $\mu'=1/(1-1/M_0)$. 
The Hamiltonian can be considered to consist of  three parts. The
first part contains 
the one-particle operators $1/(2\mu)\bbox{p}^2_i$, the  Zeeman terms
$1/(2\mu') \bbox{B}\cdot\bbox{l}_i$, the diamagnetic terms
$1/(8\mu{})(\bbox{B}\times\bbox{r}_i)^2$, the attractive Coulomb
interaction with the nucleus $-2/|\bbox{r}_i|$, and the spin Zeeman terms
$\bbox{B}\cdot\bbox{s}_i$. The second part contains the two~particle
operator (\ref{eqn:elec-elec}), which describes the repulsive Coulomb
interaction between the two 
electrons. The third operator is the so-called mass polarization
operator $H_{mp}$. It arises due to the transformation of the laboratory
coordinates to the internal coordinates, which are relative to the nucleus.
The reader should note that the Hamiltonian (\ref{eqn:hamiltonian}) has
the same good quantum numbers 
as the Hamiltonian for infinite nuclear mass: the total spin
$\bbox{S}^2$, the z~component $S_z$ of the total spin, the magnetic
quantum number $M$ and the total spatial z~parity $\Pi_z$ 
(parity is not an independent symmetry, it can be deduced from these
corresponding symmetry operations). In the following we will denote
the states by $\nu^{2S+1}M^{\Pi_z}$, where $2S+1$ is the spin
multiplicity and $\nu=1,2,3\ldots$ denotes the degree of excitation
within a given subspace.

If we compare $H(M_0,\gamma)$ in Eq.\ (\ref{eqn:hamiltonian}) to the
electronic Hamiltonian of helium in the infinite 
nuclear mass frame \cite{Becken:1999_1} $H(\infty,\gamma)$,  we  observe
two different kinds of corrections due to the finite nuclear mass. The
first is due to the 
occurrence of the reduced masses $\mu$ and $\mu'$ in $H_{rm}$ which are
the so called
normal  mass corrections. The spectrum of the Hamiltonian $H_{rm}$ which 
contains exclusively these normal mass corrections can be related to
the spectrum of 
the Hamiltonian $H(\infty,\gamma')$ at a different field
strength $\gamma'$, via a unitarian transformation and an additional trivial
energy shift (see section \ref{sec:finite_mass} and also Refs. \cite{Ruder:1994_1,Becken:1999_1,Pavlov-Verevkin:1980_1}).
The second type of finite nuclear mass effects is due to the mass polarization
operators $H_{mp}$. We will call these specific finite mass
corrections. They are by no means trivial and are
related to the correlation of the electrons (the corresponding
operators contain one-particle
operators of both  electrons). If the electrons behave
in a correlated way, the specific nuclear mass effects are enhanced.
Note that these operators are a consequence of the transformation from
the laboratory to
our non-inertial system thereby eliminating the CM degrees of freedom.

\subsection{Technical remarks}

Some comments concerning our computational approach are in order. Its
basic ingredient is an 
 anisotropic Gaussian basis set, which was put forward by
Schmelcher and Cederbaum \cite{Schmelcher:1988_1}. This one-particle basis
is sufficiently flexible to to describe finite electronic systems for
any  field strength and was
successfully applied
to several atoms and molecules in  magnetic fields
\cite{Becken:1999_1,Becken:2000_1,Becken:2001_1,Kappes:1996_1,Detmer:1997_1,Detmer:1998_1,Al-Hujaj:2000_1}.

This basis set is not only flexible: in the case of atoms all matrix
element can be 
calculated analytically
and in particular evaluated efficiently
\cite{Becken:1999_1,Becken:2000_1}. The price, which 
has to be paid, is that for each
field strength and each symmetry, the basis set has to be
(nonlinearly) optimized. This is achieved by variationally computing the
eigenfunctions of the corresponding
 one-particle problems (H, He$^+$, etc). 
This task is time consuming and needs  experience, because
the starting values for the nonlinear variational parameters have to
be chosen carefully in oder to obtain
well-converged results.
In recent times however this has become
more comfortable and was speeded up by new optimization algorithms and
tools, which allow an almost automatic construction of  a basis
set. The implementation of these tools still needs a proper selection
of the
coefficients and an evaluation of the outcome of the
optimization. The latter is  of particular importance for the field regime
addressed in the present
work for which small changes in the basis set
can have a considerable effect on the results of the full problem,
i.e. the eigenenergies and eigenfunctions of $H(M_0,\gamma)$ in
Eq.\ (\ref{eqn:hamiltonian}). 
A  one particle basis set typically consists of 200--400 basis
functions, which combine  3000 to 5000 two particle
configurations. The full CI approach leads then to a generalized
eigenvalue problem. Numerical problems arise in this generalized
eigenvalue problem if near  linear dependencies
in the basis set arise. These dependencies cannot fully be avoided, but
the resulting numerical instabilities can be removed by a cutoff of the small
eigenvalues of the corresponding overlap matrix.

The most CPU time consuming part in the above approach, is the evaluation
of the electron-electron matrix elements. Although  analytical
formulas for all 
matrix elements are available, it is important to have efficient algorithms for
their evaluation, since  the matrix elements for the
electron-electron Coulomb interaction are by no means trivial.
A detailed and sophisticated analysis of their analytical representation
is crucial. In the simplest form it contains multiple sums over  
hypergeometric functions \cite{Becken:1999_1,Becken:2000_1}. For the
evaluation of the hypergeometric function 
all possible analytical continuation formulas have been worked out
(see Ref. \cite{Becken:2000_2}) 
and made it possible to reduce the time
for the calculation of the matrix elements by a factor of 50
compared to a
straightforward implementation. We emphasize that the present investigation
would have been impossible without this efficient implementation of
the evaluation of the matrix elements.

\subsection{Threshold}
\label{subsec:threshold}

The result of the solution of the generalized eigenvalue problem  are
the total energies of the 
system. These energies are however not of primary interest: they
increase almost  linearly with the magnetic field
strength due to the raise of the kinetic energy in the presence of the
field. The relevant energies are e.g. the transition and ionization
energies which are for the high field regime addressed here by several orders of magnitude smaller than the
total energies and therefore `hidden' in the total energies.
To obtain the ionization energies we need to know the energy of the
He$^+$ ion taking into account the finite nuclear mass in the magnetic
field. These energies for He$^+$ are unfortunately not known
exactly. In contrast to this the energies for He$^+$ assuming an
infinite nuclear mass can be calculated from the well-known energies
for the hydrogen atom with fixed nucleus in strong magnetic fields (see
e.g. Ref.\cite{Kravchenko:1996_1}) via the corresponding scaling
relations (see for example \cite{Ruder:1994_1}).
However the energies of the He$^+$ ion with finite nuclear mass cannot
be extracted from results for the hydrogen atom (with finite or
infinite nuclear mass) because of the unique coupling between the
collective and electronic motion which has to be taken into
account. This coupling inherently mixes the collective and electronic
motion and 
the exact ground state therefore combines both motions. The moving
He$^+$--ion in a strong magnetic field is a  problem
which has not been solved numerically exactly in the literature. As a
result the exact threshold 
energy for He$^+$~+~e$^-$ is not known. Using the threshold energy of
He$^+$~+~e$^-$ for fixed nucleus E$^{th}_{fn}$ provides a wrong
description of the threshold energies in superstrong fields.

The Hamiltonian for the helium positive ion reads in atomic units as
follows:
\begin{eqnarray}
  \label{eq:hamil_he+}
  H_0&=&H_a+H_b+H_c\\
  H_a &=&\frac{1}{2 (M_0+1)}
  \left(\bbox{P}-\frac{1}{2}\bbox{B}\times\bbox{R}\right)^2\\
  H_b &=&
  \frac{M_0+2}{(M_0+1)^2}\left(\bbox{P}-\frac{1}{2}\bbox{B}\times\bbox{R}\right)\cdot\left(\bbox{B}\times
    \bbox{r}\right)\\
  H_c &=& \frac{1}{2\mu}\bbox{p}^2 +
  \frac{1}{2}\left(\omega_1-\frac{\omega_2}{M_0}\right)\bbox{B}\cdot
  \bbox{l} + \frac{1}{8}\left(\omega_1^2+\frac{\omega_2^2}{M_0}\right) \left(\bbox{B}\times\bbox{r}\right)^2.
\end{eqnarray}
Here $H_a$ describes the collective motion of the ion as  a free
particle with charge 
1 and mass $M_0+1$ moving in a magnetic field ($\bbox{R}$ and $\bbox{P}$
are the center of mass coordinate and momentum, respectively). The operator $H_b$ 
couples electronic and collective 
motion. The operator $H_c$ is the electronic part of the He$^+$
Hamiltonian with $\omega_1=1+1/(M_0+1)^2$ and $\omega_2=1+(2M_0+1)/(M_0+1)^2$.

Although the exact threshold energy for He$^+$~+~e$^-$ is not available
some approximations to it have been calculated in the literature. 
 One of those approximations E$^{th}_{zpmc}$
ignores the coupling between the collective and electronic motion $H_b$. The 
corresponding threshold values consist of the sum of eigenenergies of 
the electronic Hamiltonian $H_c$ and the zero point energy
for $H_a$. The zero point energy for the collective
motion is the corresponding energy for the lowest Landau level. The energy for this
threshold is typically too high, i.e. the true ionization energies are overestimated,
since the coupling $H_b$, which is neglected in this approximation,
tends to reduce the threshold energy. All 
ionization energies shown in the present work are calculated by applying
E$^{th}_{zpmc}$.

The second approximation we will choose, 
ignores the zero point energy of the ion and the coupling term,
i.e. $H_a$ and $H_b$. Therefore, only
the eigenenergies of the mass corrected electronic Hamiltonian $H_c$ are taken
into account. This threshold 
is denoted as E$^{th}_{mc}$. It is motivated by the fact,
that for an infinitely strong 
magnetic field the energetic contributions due to the zero point
energy of $H_a$ and the coupling $H_b$ exactly cancel.
A third alternative threshold energy E$^{th}_{ad}$ is obtained by employing the
adiabatic expansion approach presented in \cite{Bezchastnov:1998_1}.
 Since this approximation only takes into account the lowest
Landau energy for the CM motion, it becomes increasingly accurate for increasing energetic separation between
the Landau levels (and therefore increasing magnetic field strength).
 For field strengths below $2000$ a.u. this
approximation is not reliable.  It can be observed, that the threshold
E$_{ad}^{th}$ approaches the
threshold energy E$_{mc}^{th}$ in the limit of high
fields. E$_{ad}^{th}$ is expected to be very accurate for sufficient
high field strengths.

\section{The finite nuclear mass effects}
\label{sec:finite_mass}

 There is a transformation, which connects the 
spectrum of the infinite nuclear mass Hamiltonian $H(\infty,\gamma)$
with the spectrum of the 
Hamiltonian H$_{rm}$ \cite{Becken:1999_1}. This transformation reads
as follows: 
\begin{equation}
  U\,H_{rm}\,U^{-1}=\mu
  H(\infty,\frac{\gamma}{\mu^2})-\frac{\gamma}{M_0} \sum_i(\bbox{l}_{z_i}+\bbox{s}_{z_i}).\label{eq:trans}
\end{equation}
Here $U$ denotes a unitarian transformation, which transforms
$\bbox{r}\rightarrow\bbox{r}/\mu$ and
$\bbox{p}\rightarrow\bbox{p}\mu$. The second term on the right hand
side of Eq. (\ref{eq:trans}) represents a field dependent trivial
energy shift, since the total z~component of the spin and the total
z~component of the orbital angular momentum are conserved quantities.

From equation (\ref{eq:trans}) it can be seen, that the leading mass
correction is 
of the order of $\gamma/M_0$. For the 
states with
magnetic quantum number $M<0$, this
correction is positive. 
This means that the corresponding energy levels are shifted 
i.e. increase linearly with $\gamma$ and eventually pass the
ionization threshold. The latter implies that the corresponding
bound electronic state becomes ionized. This shift depends exclusively on
the magnetic quantum number $M$, the total z~component of the spin
$S_z$, the magnetic field strength $\gamma$, and the nuclear mass M$_0$.

\begin{figure}[htbp]
  \begin{center}
    \epsfig{file=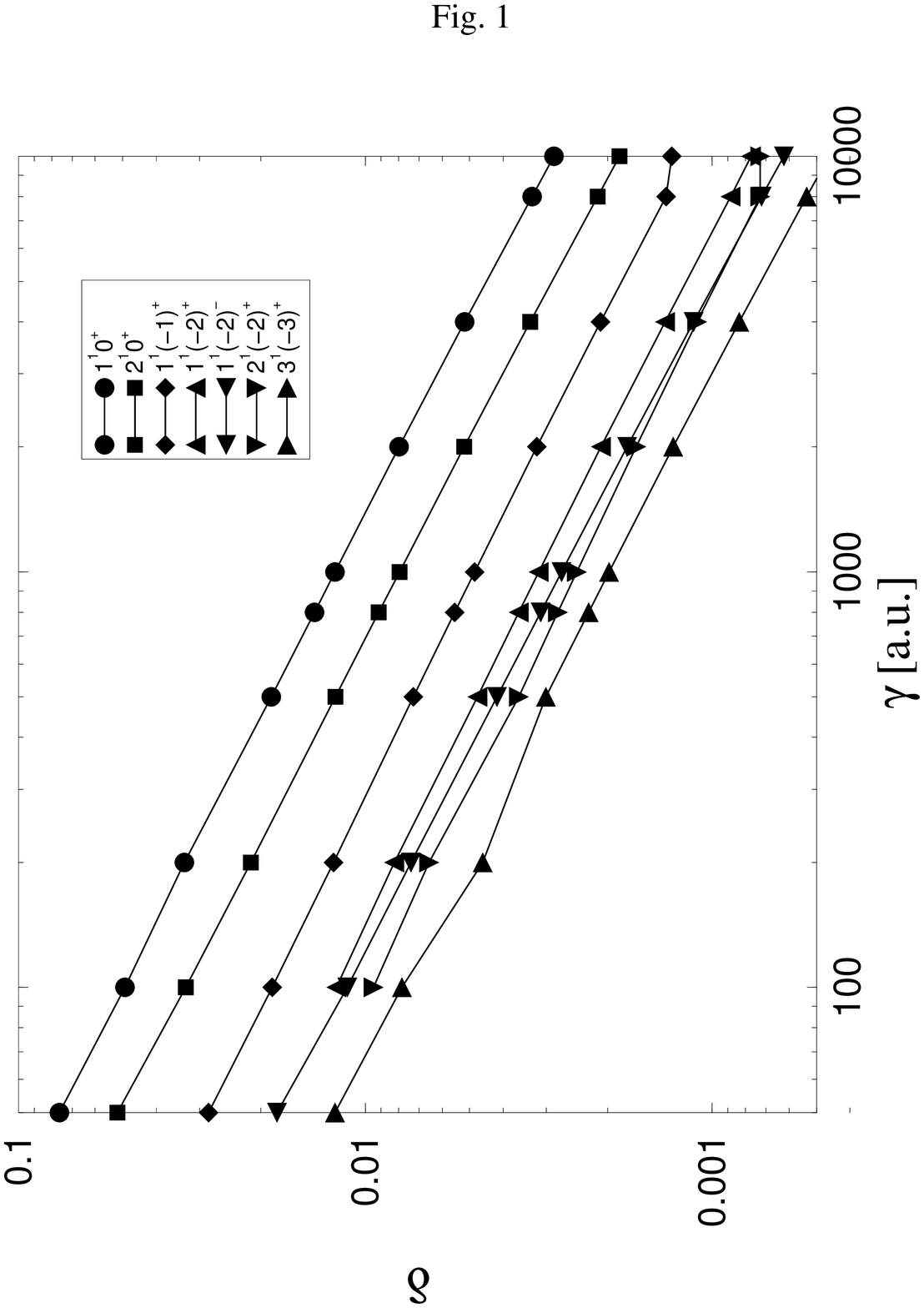,angle=-90,width=9cm}    
    \caption{The parameter $\delta$ as a function of the field
      strength for singlet states of various symmetries. It reflects
      the  state dependent normal finite mass corrections due to the
      Coulomb interaction. For details see text.}
    \label{fig:Rel_Corr}
  \end{center}
\end{figure}

The energy difference
$\Delta E_{rm}$=E$_{rm}(\gamma)-$E$(\infty,\gamma)$, where $E_{rm}$
denotes the eigenvalues of $H_{rm}$ and $E_(\infty,\gamma)$ the
eigenvalues of $H(\infty,\gamma)$, can be related to
the energy difference $\Delta E^e_{rm}$ of the eigenvalues of the
Hamiltonian H$_f^e$ and H$_{rm}^e$. The operator H$_f^e$ describes two free
noninteracting electrons
\begin{equation}
  \label{eq:hamil_free_elect}
  H_{f}^e = \sum_{i=1}^2\frac{1}{2}\bbox{p}_i^2+\frac{\gamma}{2}l_{z_i}+\frac{\gamma^2}{8}\rho_i^2
\end{equation}
whereas H$_{rm}^e$ refers to the corresponding `artificial'
Hamiltonian with reduced masses
\begin{equation}
  \label{eq:hamil_artificial}
  H_{rm}^e =  \sum_{i=1}^2\frac{1}{2\mu}\bbox{p}_i^2+\frac{\gamma}{2\mu'}l_{z_i}+\frac{\gamma^2}{8\mu}\rho_i^2.
\end{equation}
For the energetically lowest Landau level with negative magnetic
quantum number $M$ and vanishing momentum in z~direction we have
\begin{equation}
  \label{eq:energy_diff}
  \Delta E^e_{rm}=\frac{\gamma}{M_0}\left(1+|M|\right).
\end{equation}
It cannot be expected, that the normal mass corrections of the helium
atom $\Delta E_{rm}^e$ follow exactly equation (\ref{eq:energy_diff}),
because the Hamiltonian H$_{rm}$ contains the interaction between the
two electrons and of the electrons with the nucleus. However it is
suggestive to introduce a parameter $\delta(\gamma)$ such that 
\begin{displaymath}
 \Delta
 E_{rm}=E_{rm}(\gamma)-E(\infty,\gamma)=\Delta
 E_{rm}^e(\gamma)\left\{1+\delta(\gamma)\right\}.
\end{displaymath}
For the states and field strengths investigated in the present work we have
$\delta(\gamma)\ll 1$. Therefore $\delta$ can be treated as a
correction term.  Since this correction is due to the Coulomb
interaction  it is state
dependent. In Fig. \ref{fig:Rel_Corr} the quantity $\delta$ is shown as a
function of the magnetic field strength $\gamma$ for a few selected
singlet states belonging to several different symmetries. It can be
seen, that $\delta(\gamma)$, for all states considered, follows a
 power law
$\delta(\gamma)=C\cdot \gamma^{-\lambda}$, where the exponent
$\lambda\approx0.62$ does not  depend on the state. However the
corresponding proportionality constant $C$ varies over nearly one
order of magnitude for the different states.

To understand more on the behavior of the quantity $\delta$  we expand the first
part on the right hand side of Eq. (\ref{eq:trans}) in powers of
$1/M_0$. Omitting the spin part we obtain:
\begin{eqnarray}
  \label{eq:devE}
\mu E(\infty, \frac{\gamma}{\mu^2})=  E_{rm}(\gamma)+\frac{\gamma M}{M_0} &=&
  \mu\left\{E(\infty,\gamma)+\frac{2 \gamma
  E'(\infty,\gamma)}{M_0}+O(\frac{1}{M_0^2})\right\}.\\
&\approx& E(\infty,\gamma)-\frac{E(\infty,\gamma)}{M_0}+2\gamma \frac{E'(\infty,\gamma)}{M_0}.
\end{eqnarray}
Here the prime indicates the derivative with respect to the field
strength. Now $\Delta E_{rm}$ and consequently $\delta$  can be
expressed in terms of the eigenenergies 
of H($\infty,\gamma$):
\begin{equation}
\label{eq:delta}
  \delta= \frac{-E(\infty,\gamma)+2\gamma E'(\infty,\gamma)+\gamma|M|}{\gamma(1+|M|)}-1.
\end{equation}
Now the state dependency of $\delta$ can be understood as the
dependence on the derivative of the eigenenergies $E(\infty,\gamma{})$
with respect to the field strength. We emphasize, that the
quantities $E(\infty,\gamma)$ and $\gamma E'(\infty,\gamma)$ in
Eq. (\ref{eq:delta}) are almost equal and therefore approximately
cancel in the superstrong field regime. Thus $\delta(\gamma)$ can be
approximated by
$\delta(\gamma)\approx(E'(\infty,\gamma)+|M|)/(1+|M|)-1$.

\begin{figure}[htbp]
  \begin{minipage}[c]{16cm}
\epsfig{file=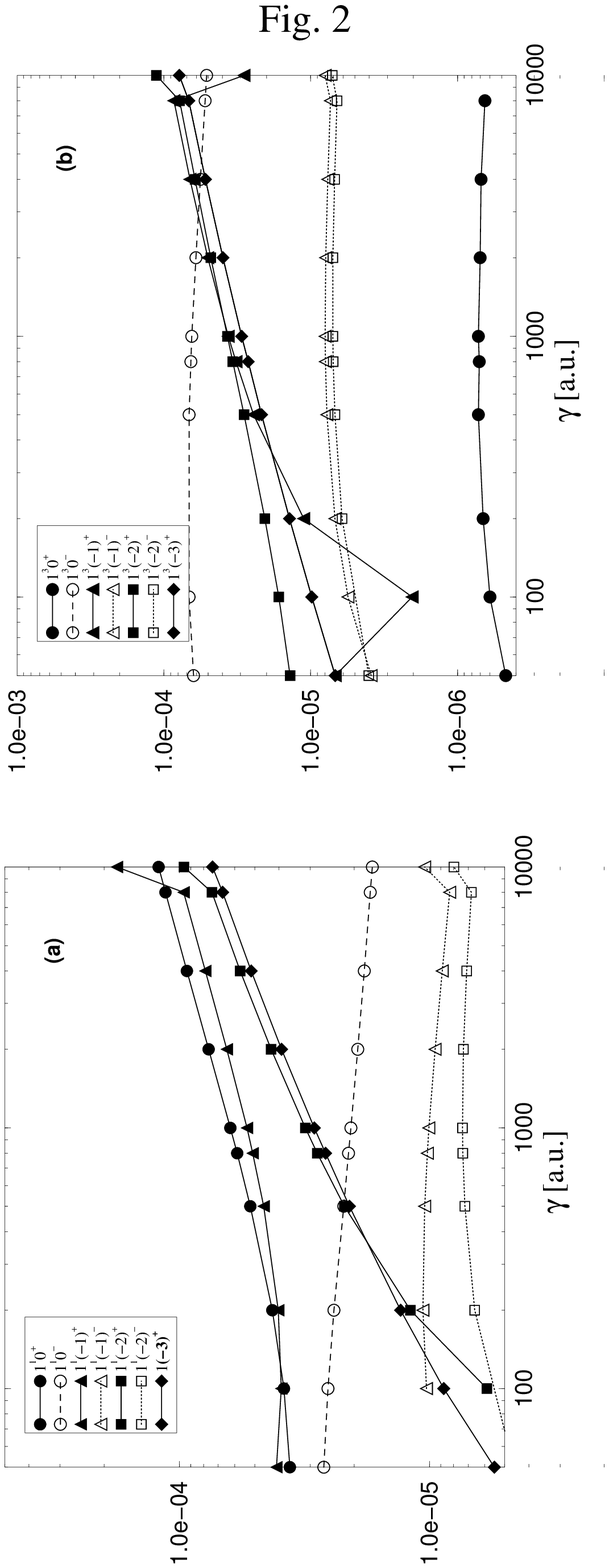,angle=-90,width=15cm}
\end{minipage}
\caption{The absolute values of the mass polarization energies
    $|E_{mp}|$=$|E(M_0,\gamma)-E_{rm}(\gamma)|$ for selected singlet
    (a) and triplet (b) states as a function of the magnetic field
    strength $\gamma$.}
  \label{fig:mass_pol_corr}
\end{figure}

Fig.\ \ref{fig:mass_pol_corr} (a) illustrates the mass polarization
energies $|E_{mp}|=|E(M_0,\gamma)-E_{rm}(\gamma)|$ for the energetically
lowest singlet states
and Fig.\ \ref{fig:mass_pol_corr} (b) for the corresponding triplet
states. Here $E(M_0,\gamma)$ denotes the eigenenergies of
$H(M_0,\gamma)$. First we  
observe, that the absolute values of $E_{mp}$ are very small: For
$\gamma=10^4$ they are at least eight orders of magnitude smaller than
the corresponding total energies. 
They are also small  compared to the normal finite mass corrections
$\Delta E_{rm}$. For $\gamma=10^4$ $E_{mp}$ is at least four orders of
magnitude smaller  the normal finite mass corrections.
 Opposite to the quantity $\Delta
E_{rm}$ the behavior of $|E_{mp}|$ depends strongly on the state.  For
the states 
$1^10^+$,$1^1(-1)^+$,$1^1(-2)^+$,$1^1 (-3)^+$ the quantity $|E_{mp}|$
increases. 
These states contain so-called magnetically tightly bound
orbitals \cite{Ruder:1994_1,Loudon:1959_1,Ivanov:1998_1} and therefore  their
ionization energy diverges
logarithmically in the limit  $\gamma\rightarrow\infty$. An
increase of the influence of $|E_{mp}|$ can also be  observed in
Fig.\ \ref{fig:mass_pol_corr} (b) for the states
$1^3(-1)^+$,$1^3(-2)^+$ and $1^3(-3)^+$, which represent also 
magnetically tightly bound states.  
For the remaining states $1^10^-$, $1^1(-1)^-$ and the corresponding
triplet states as well for $1^30^+$ $|E_{mp}|$ remains almost constant
as a function of the magnetic field strength.
This effect
can be easily understood: For the magnetically tightly bound states the
electrons are close to each other in a relatively narrow region of
space and therefore electron correlation is important. 
 The mass polarization operator is
sensitive to  electronic correlation, due to the fact that
it  contains operators of both electrons.
The sign of $E_{mp}$ is not shown in Fig.\ \ref{fig:mass_pol_corr}. It
is positive for the states related to the above mentioned tightly
bound states and negative for the others. The only exception is
 the state $1^30^+$, which does not belong to the tightly
bound states, but nevertheless $E_{mp}$ has a positive sign.

\section{Results}
\label{sec:results}

In the following we will present our results for the ionization
energies and transition wavelengths of the helium atom for magnetic
fields ranging from $100$~au to $10000$~au. These investigations have
been performed for the magnetic quantum numbers $M=0,-1,-2,-3$, singlet and triplet states and positive and negative
z~parity. Only for the magnetic quantum number $M=-3$ exclusively
positive z~parity  states have been studied. For most symmetry subspaces we
investigated 6 excited states.

\subsection{Ionization energies}

According to the above the reader should keep in mind, that the exact
energy of the ionization threshold  is not known, and therefore the ionization
energies are not known accurately. However the ionization energies
calculated by using different approximative threshold energies
$E^{th}_{fn}$, $E^{th}_{mc}$, $E^{th}_{zpmc}$, $E^{th}_{ad}$\ show
the same trend: the number of bound states of the helium atom becomes
finite for superstrong magnetic fields in contrast to the situation
without a magnetic field, or in the limit of an infinitely heavy
nucleus,  where the helium atom has an infinite number
of bound states. Only the so-called magnetically tightly bound states
are bound within the complete regime of field strengths and quantum
numbers considered in
the present work. We consider in the following the quantity
$E_{ion}=E^{th}_{zpmc}-E(M_0,\gamma)$ as a function of the field strength
together with the above mentioned approximations for the threshold.

\begin{figure}[htbp]
  \begin{center}
    \epsfig{file=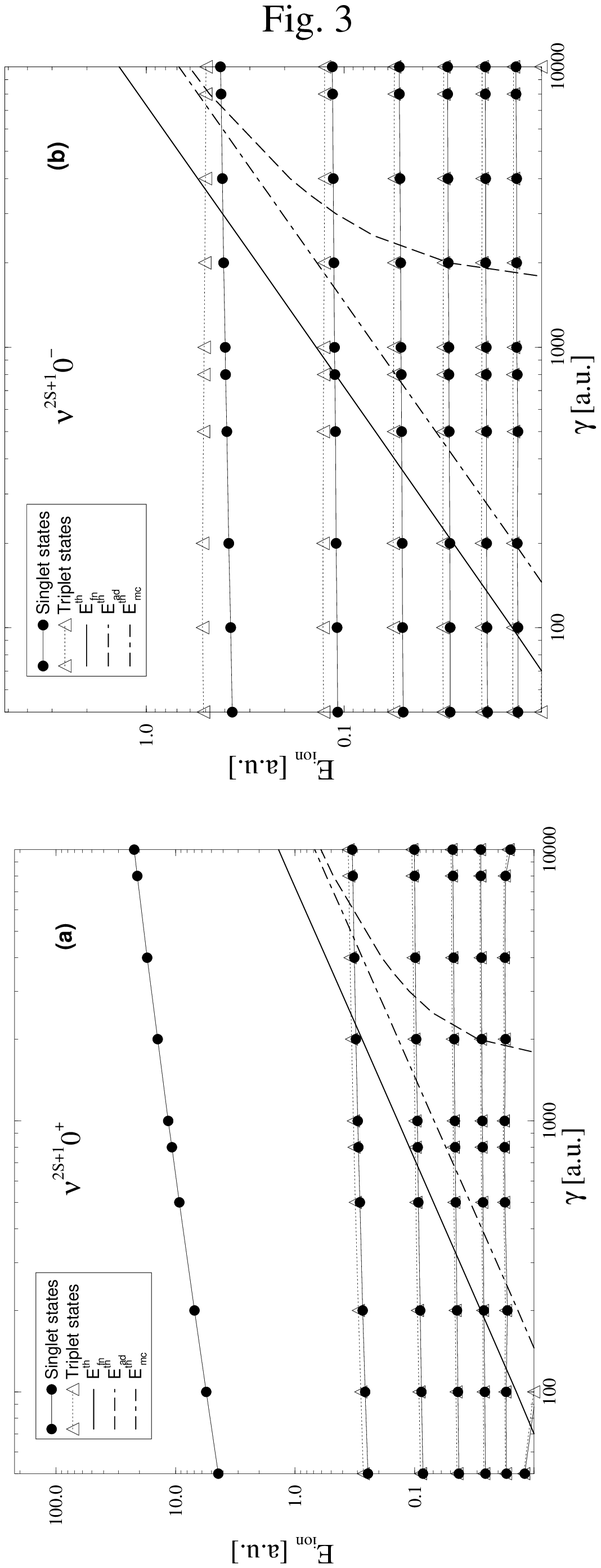,angle=-90,width=14cm}
    \caption{Ionization energies $E_{ion}=E^{th}_{zpmc}-E(M_0,\gamma)$ of
      the helium atom 
      for the $\nu 
      ^{2S+1}0^{\pi_z}$ states. (a)  The ionization energies
      for the states with positive z~parity. (b) ionization energies for the
      corresponding states with negative z~parity. Note that the
      threshold $E^{th}_{zpmc}$ overestimates the ionization
      energies. Different approximations to the exact threshold
      $E^{th}_{mc}$, $E^{th}_{ad}$ and the fixed nucleus ionization
      threshold 
      $E^{th}_{fn}$ are given.}
    \label{fig:he_bind_m0}
  \end{center}
\end{figure}

Fig.\ \ref{fig:he_bind_m0} (a) shows $E_{ion}$ for the six
energetically lowest states of zero magnetic quantum number and
positive z~parity. The $1^10^+$ state is the most tightly bound
state. In strong magnetic fields it represents however not the ground
state of the atom, because energetically low-lying states are fully
spin-polarized in the high field regime. The ground state is given by
the $1^3(-1)^+$ state, which is also a tightly bound
state. Fig.\ \ref{fig:he_bind_m0} (a) shows, that $E_{ion}$ increases
for the tightly bound state $1^10^+$, but remains approximately constant for
all other states as a function of the magnetic field strength. Furthermore
we observe that 
 all states $\nu
^10^+$ with $\nu>1$ as well as the corresponding triplet states pass
the threshold energies being either $E^{th}_{ad}$ or $E^{th}_{mc}$
with increasing field strength. The only remaining bound state for
$\gamma=10^4$ au is the $1^10^+$ state.

In Fig.\ \ref{fig:he_bind_m0} (b) the corresponding quantity for states
with $M=0$ and negative z~parity are shown. For these states $E_{ion}$
as a function of $\gamma$ varies only to a very minor extent, which is
due to the 
fact, that none of these states is a tightly bound one, and none
of these states remains bound when $\gamma$ approaches $10^4$~au. For
the triplet states with negative z~parity the quantity $E_{ion}$
decreases slightly, for $\gamma>100$~au.
 This is not due to the finite mass effect, but can be also
observed for the quantity $E_{fn}^{th}-E(\infty,\gamma)$, whereas this
quantity increases monotonically for all 
other states investigated in the present work. The reason is the
complicated interplay between 
correlation, which tends to increase ionization energies and Coulomb
repulsion, which tends to decrease it. For the states
with $M=0$ the electrons are confined in a very small domain of space,
which increases correlation as well as the Coulomb repulsion. On the
other hand for triplet states the electrons are separated, because the
wavefunction is antisymmetric, which reduces both effects. For the
$\nu ^30^-$ states the increase of correlation energy is smaller than
the increase of the Coulomb repulsion energy.

\begin{figure}[htbp]
  \begin{center}
    \epsfig{file=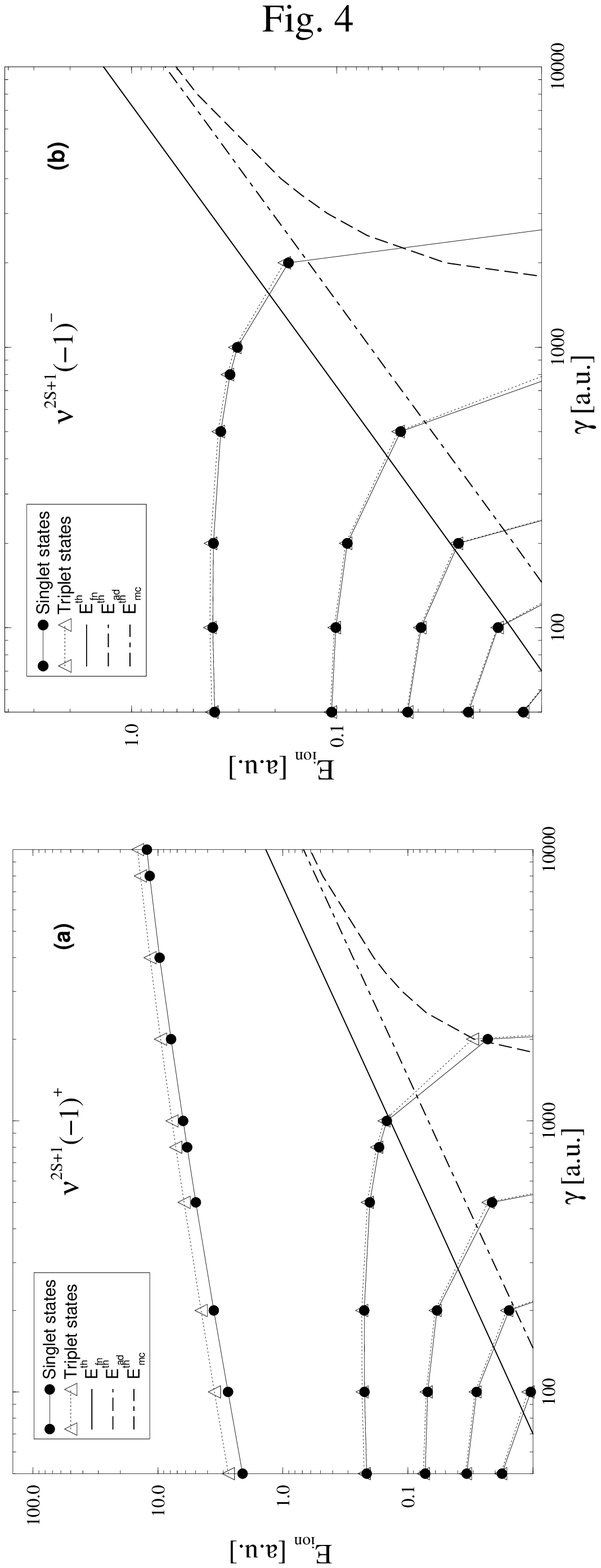,angle=-90,width=14cm}
    \caption{Ionization energies $E_{ion}=E^{th}_{zpmc}-E(M_0,\gamma)$ of
      the helium atom for the electronic states with $M=-1$ as a
      function of the magnetic field strength $\gamma$. (a) $E_{ion}$
      for positive z~parity states.
      (b) $E_{ion}$  for negative
      z~parity states.} 
    \label{fig:he_bind_m1}
  \end{center}
\end{figure}

In figure \ref{fig:he_bind_m1} (a) $E_{ion}$ is shown for the states
$\nu ^{2S+1}(-1)^+$.  For the magnetically tightly bound
states $1^1(-1)^+$ and $1^3(-1)^+$ the ionization energy remains
positive, i.e. these states are bound in the complete regime $\gamma<10^4$~au.
For  higher excited states, i.e. $\nu^{2S+1}(-1)^+$ with $\nu>1$
the energy $E(M_0,\gamma)$ becomes even larger than $E^{th}_{zpmc}$
and therefore $E_{ion}$ decreases strongly on the logarithmic
scale. To understand this we review Eq. (\ref{eq:trans}): The dominant
term on the right hand side is of the form $-M\gamma/M_0$ (the spin
part  does not affect the ionization
energies). Therefore  $E(M_0,\gamma)$ for states with $M<0$ 
raises about this amount and will pass the threshold at lower field
 strength than their counterparts with $M=0$.

In Fig.\ \ref{fig:he_bind_m1} (b) we show our results for $E_{ion}$ for
$M=-1$ and negative z~parity. A similar behavior to that of  the states $\nu ^{2S+1}(-1)^+$,
with $\nu>1$ is observed: The ionization energy as a function of the magnetic
field strength decreases rapidly. This is due to the fact, as
mentioned above, that the finite mass corrections force these states
to pass even the $E^{th}_{zpmc}$ threshold and the states become
unbound. For negative z~parity  tightly bound states do not exist.

\begin{figure}[htbp]
  \begin{center}
    \epsfig{file=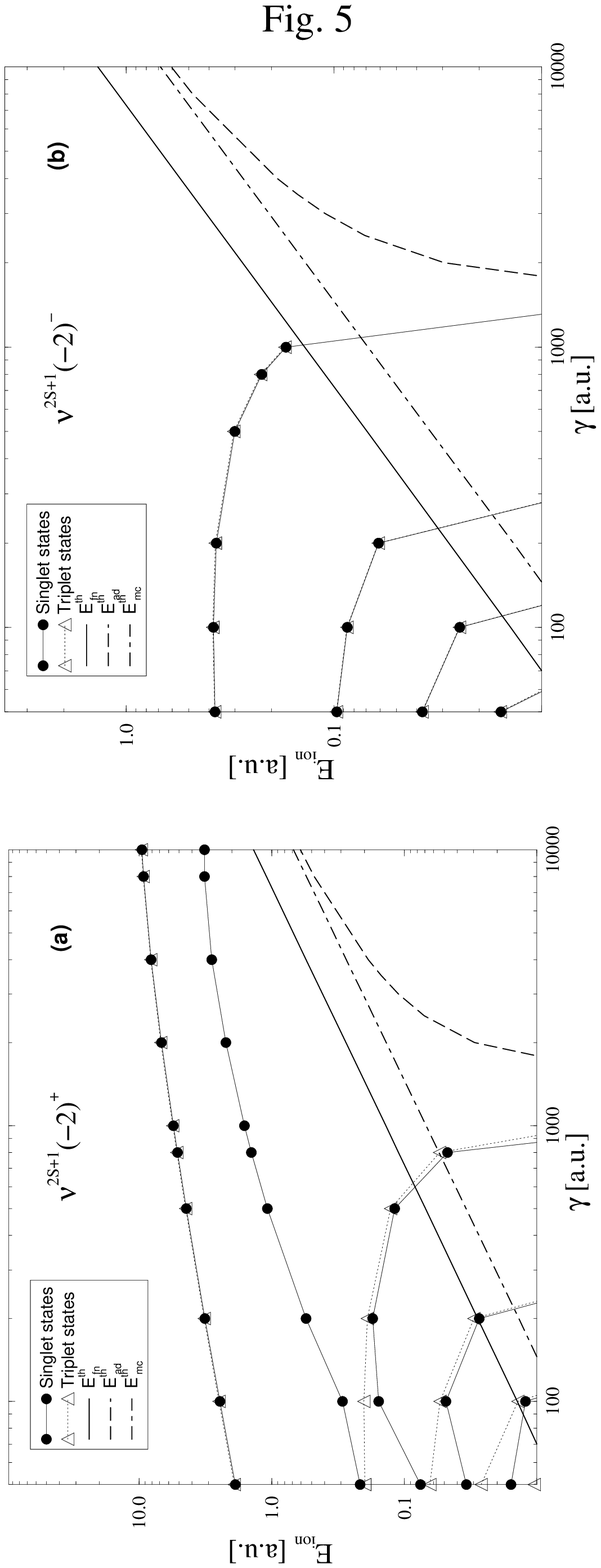,angle=-90,width=14cm}
    \caption{$E_{ion}$ for the electronic states
      $\nu^{2S+1}(-2)^{\Pi_z}$. (a) For positive z~parity  we encounter an
      avoided crossing, which gives rise to a significant  increase of  $E_{ion}$
       for the $2^1(-2)^+$ state. (b) $E_{ion}$ for the corresponding
      negative z~parity states.} 
    \label{fig:he_bind_m2}
  \end{center}
\end{figure}

In Fig.\ \ref{fig:he_bind_m2} (a) we present our results for $E_{ion}$ 
for the energetically lowest  singlet and triplet states with $M=-2$ and
positive z~parity. Similar to Fig. \ref{fig:he_bind_m1} (a) the quantity
$E_{ion}$ rises for the two energetically lowest singlet and triplet states from
$1.8$ au to approximately 10 au. These two states belong to the
magnetically tightly bound states. However in contrast to $E_{ion}$
for the $2^{2S+1}(-1)^+$ states in
Fig. \ref{fig:he_bind_m1} (a) $E_{ion}$ for the 
state $2^1(-2)^+$, which is the first excited singlet state of this
symmetry,  does also rise and stays bound within the complete regime of field
strength considered here. This is a remarkable feature
since the influence of the finite mass effects for this state is even
bigger  than
for the states with $M=-1$. The reason for this behavior lies in the
presence of an avoided crossing 
which takes place at $\gamma\approx100$~au. It can be seen that
$E_{ion}$ for the higher excited singlet states is raised as well
and it  approaches the corresponding value for the next energetically
 higher triplet state.

The ionization energies for the negative z~parity states for $M=-2$ in
Fig.\ \ref{fig:he_bind_m2} (b) look similar to these of the $M=-1$ in
Fig. \ref{fig:he_bind_m1}
(b). Magnetically tightly bound states do not exist for $M=-2$ and
negative parity, therefore all states of this symmetry become unbound
with increasing field strength.
 As mentioned above the influence of the
finite nuclear mass increases with
increasing magnetic quantum number $|M|$, therefore the states
$\nu^{2S+1}(-2)^-$ become unbound at lower field strengths than the
corresponding states with magnetic quantum number $M=-1$.

\begin{figure}[htbp]
  \begin{center}
    \epsfig{file=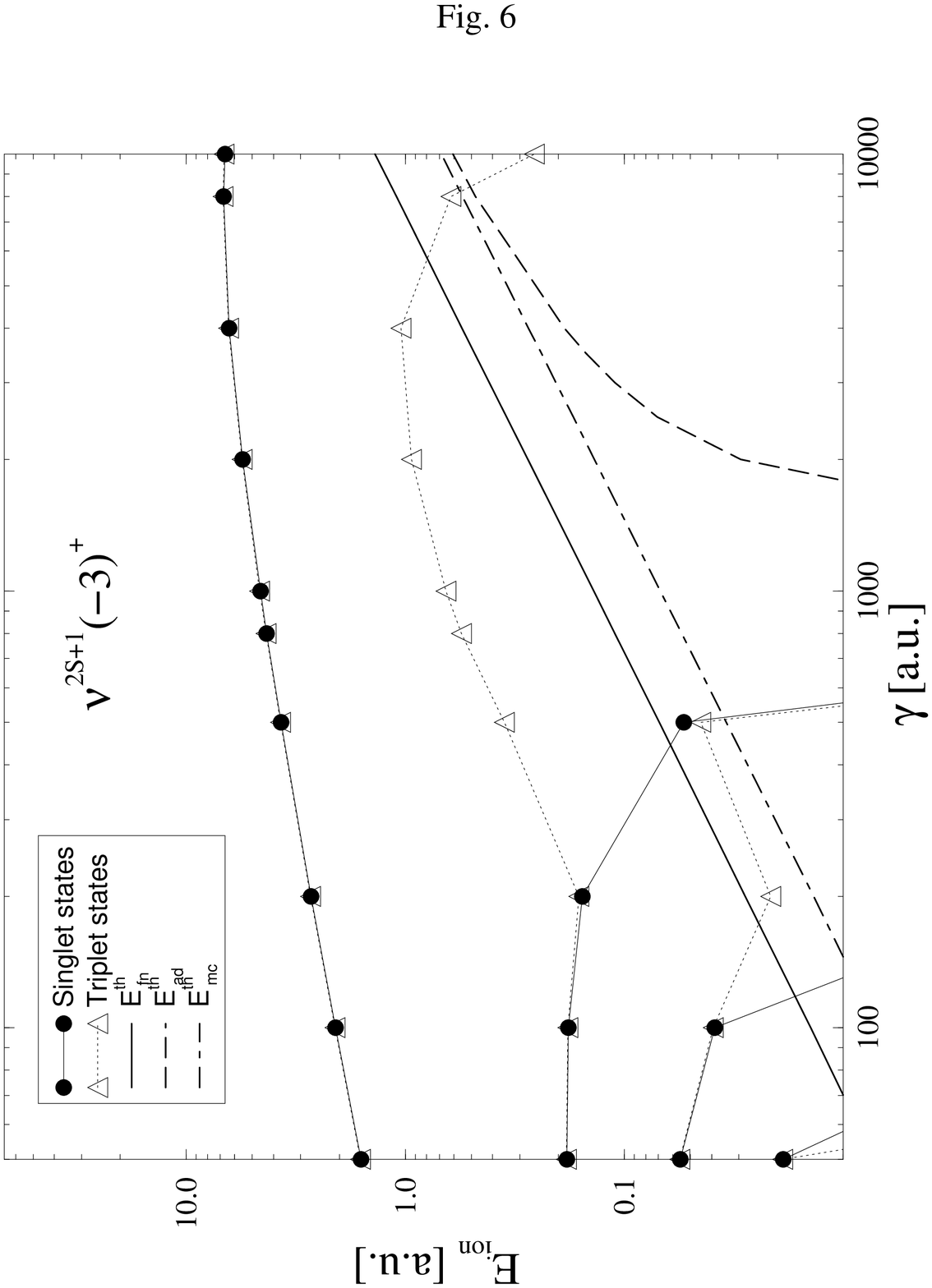,angle=-90,width=9cm}
    \caption{$E_{ion}$ for the electronic states with magnetic quantum
      number $M=-3$ and positive
      z~parity. An avoided crossing increases the ionization energy
      for the state $2^3(-3)^+$.} 
    \label{fig:he_bind_m3p}
  \end{center}
\end{figure}

In Fig.\ \ref{fig:he_bind_m3p} we encounter that $E_{ion}$ for the
 energetically lowest singlet and triplet states with $M=-3$ and
positive z~parity is positive for all field strengths, considered in the
present work. These two states belong to the magnetically tightly
bound states.
Similar to Fig.\ \ref{fig:he_bind_m2} (a) 
an avoid crossing takes place at $\gamma\approx200$~au. This can
 be more clearly seen if the quantity $E_{fn}^{th}-E(\infty,\gamma)$
 is considered. But unlike
to the case of $M=-2$, where E$_{ion}$ for the first excited singlet
state is raised, here the triplet state
$2^3(-3)^+$ is shifted to lower energies and remains therefore  bound for
relatively high  field strengths. Nevertheless, the state 
$2^3(-3)^+$ passes the ionization threshold for $\gamma$ approaching
 $10^4$~au  and 
therefore becomes unbound due to the influence of the finite
nuclear mass effects.

\subsection{Transition wavelengths}

In contrast to the ionization energies, which strongly depend on the
exact values for the threshold energies, transition wavelengths can be
calculated 
from  total energies without the knowledge of the exact threshold energy. The only
feature which remains unknown for the transition wavelengths is the
exact field strength, foe which  the particular bound-bound transition
disappears. Therefore we refer the total energy of a state to the
threshold $E_{zpmc}^{th}$ in order to decide upon its bound character
as a function of the field strength. According to the discussion
provided in Sect. \ref{subsec:threshold} the true field strength for
which the state becomes unbound is lower than the value obtained by
refering the total energies to $E_{zpmc}^{th}$.

\begin{figure}[htbp]
  \begin{center}
    \epsfig{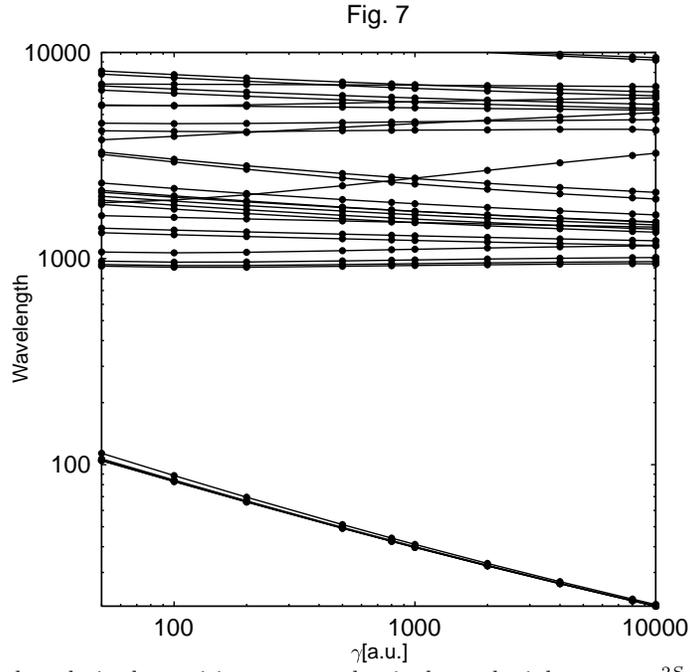}
    \caption{Wavelengths of linearly polarized transitions among the
      singlet and triplet 
      states $\nu^{2S+1} 0^+$ and $\mu^{2S+1} 0^-$ in
      {\AA}ngstr{\o}m as a function of the magnetic field strength in
      atomic units.
      The short wavelengths correspond to transitions involving the 
      most tightly bound state $1^10^+$.} 
    \label{fig:trans0+to0-}
  \end{center}
\end{figure}

\begin{figure}[htbp]
  \begin{center}
    \epsfig{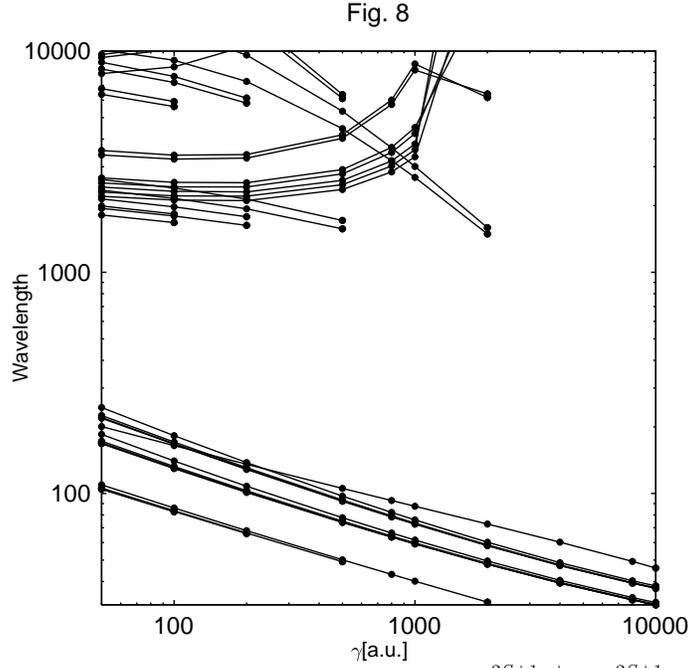}
    \caption{Transition wavelengths for the singlet and triplet
      transitions from $\nu ^{2S+1} 0^+$ to $\mu^{2S+1} (-1)^+$ in
      {\AA}ngstr{\o}m corresponding to circular polarized transitions. 
      The shortest wavelengths are due to transitions involving the states
      $1^10^+$,$1^1(-1)^+$ and $1^3(-1)^+$, which belong to the
      magnetically tightly bound states. The transitions between
      excited states are
      dominated by  finite nuclear mass effects.} 
    \label{fig:trans0+to1+}
  \end{center}
\end{figure}

\begin{figure}[htbp]
  \begin{center}
    \epsfig{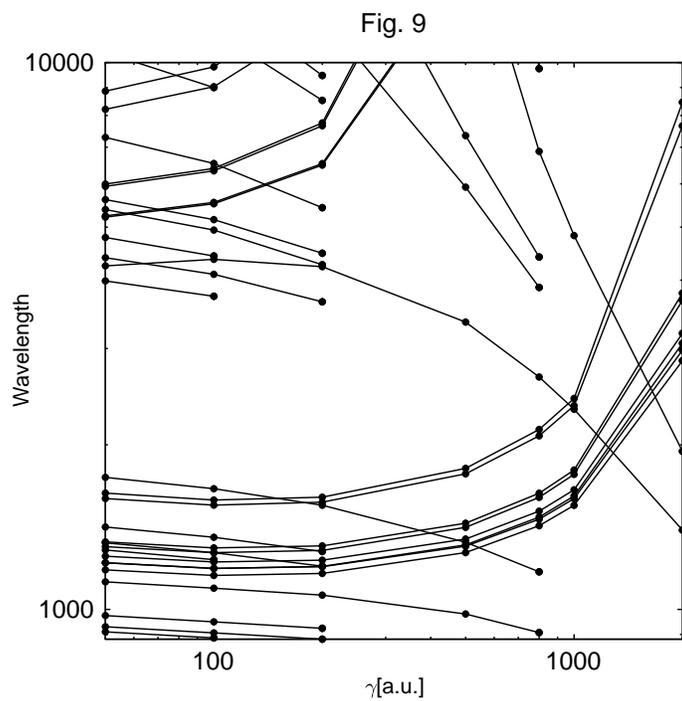}
    \caption{Transition wavelengths for the circular polarized
      transition   $\nu ^{2S+1} 0^-$ to $\mu^{2S+1} (-1)^-$ in
      {\AA}ngstr{\o}m.}
    \label{fig:trans0-to1-}
  \end{center}
\end{figure}

\begin{figure}[htbp]
  \begin{center}
    \epsfig{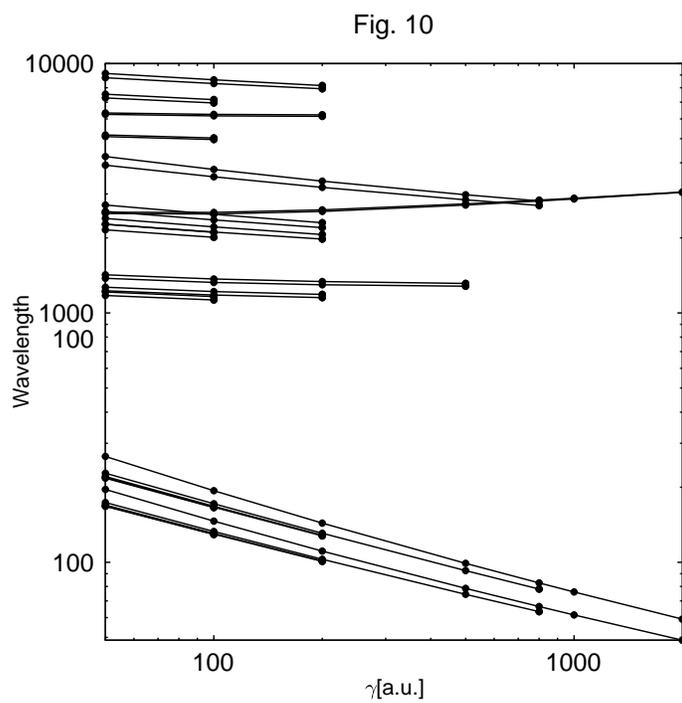}
    \caption{Transition wavelengths for singlet and triplet transitions for $\nu ^{2S+1} (-1)^+$ to $\mu^{2S+1} (-1)^-$ in {\AA}ngstr{\o}m.}
    \label{fig:trans1+to1-}
  \end{center}
\end{figure}

\begin{figure}[htbp]
  \begin{center}
    \epsfig{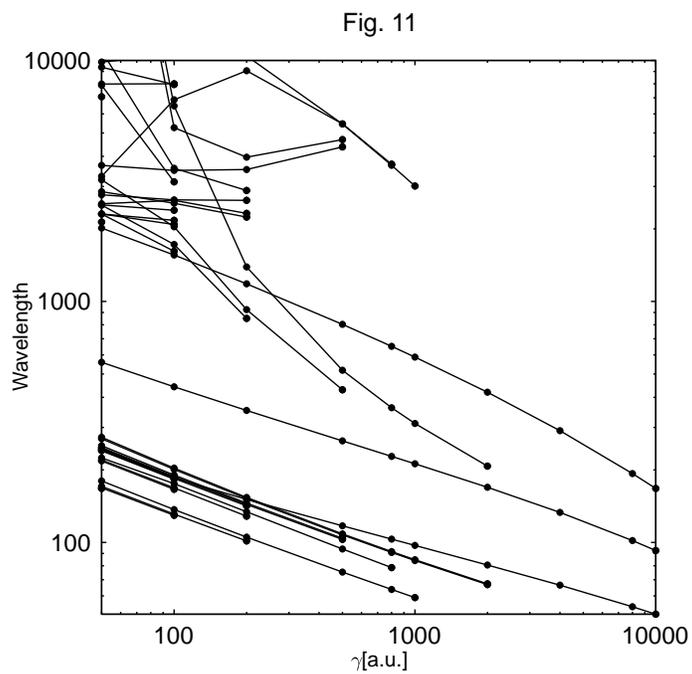}
    \caption{Transition wavelengths for the singlet and triplet transitions from $\nu^{2S+1}  (-1)^+$ to $\mu^{2S+1} (-2)^+$ in {\AA}ngstr{\o}m.}
    \label{fig:trans1+to2+}
  \end{center}
\end{figure}

\begin{figure}[htbp]
  \begin{center}
    \epsfig{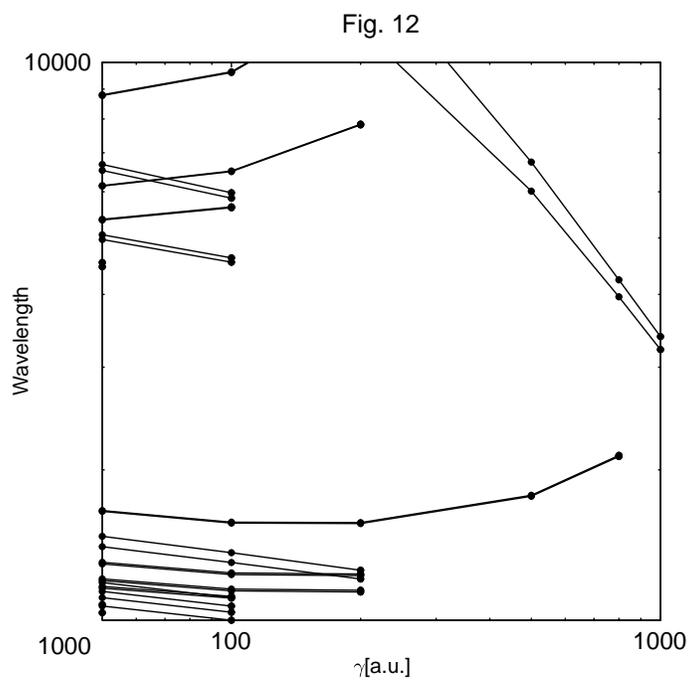}
    \caption{Transition wavelengths for singlet and triplet
      transitions from  $\nu^{2S+1}  (-1)^-$ to $\mu^{2S+1} (-2)^-$ in {\AA}ngstr{\o}m.}
    \label{fig:trans1-to2-}
  \end{center}
\end{figure}

\begin{figure}[htbp]
  \begin{center}
    \epsfig{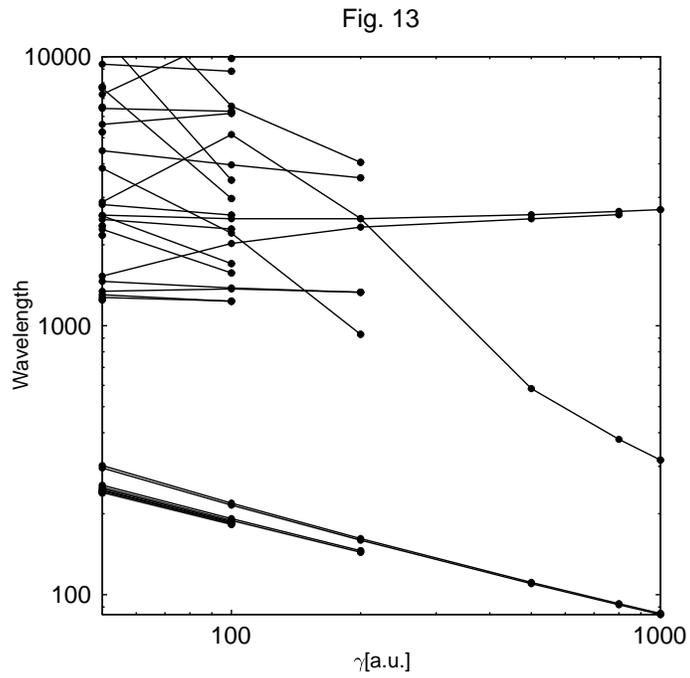}
    \caption{Transition wavelengths for singlet and triplet transitions from $\nu^{2S+1}  (-2)^+$ to $\mu^{2S+1} (-2)^-$ in {\AA}ngstr{\o}m.}
    \label{fig:trans2+to2-}
  \end{center}
\end{figure}

\begin{figure}[htbp]
  \begin{center}
    \epsfig{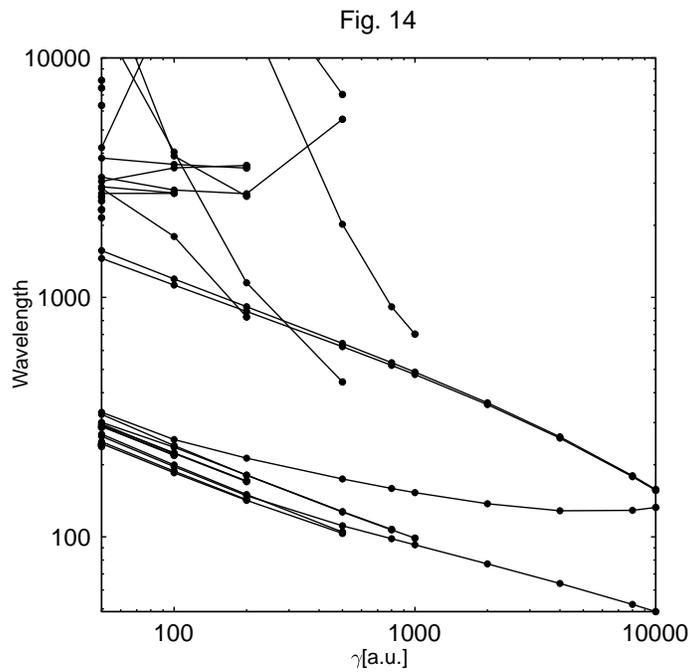}
    \caption{Transition wavelengths for singlet and triplet transitions from $\nu^{2S+1}  (-2)^+$ to $\mu^{2S+1} (-3)^+$ in {\AA}ngstr{\o}m.}
    \label{fig:trans2+to3+}
  \end{center}
\end{figure}

Some general remarks on our results for the transition wavelengths
presented in Figs. \ref{fig:trans0+to0-}~--~\ref{fig:trans2+to3+} are
in order.
Linearly polarized transitions show the general feature, that there are two
separated parts of the spectrum. A few transition wavelengths  become shorter, following
approximately a power law as a function of the field strength. Other
transition wavelengths, which are much longer, stay almost constant 
 as a function of the magnetic field strength. The short
wavelengths correspond to  transitions involving  the magnetically
tightly bound 
states. The other lines correspond to transitions between
higher excited states.

For circular polarized transitions the appearance of the spectrum of
wavelengths is
different.  Only for circular polarized transitions which
involve states with positive z~parity  a  separated bundle of
short wavelengths is present. This is again due to the presence of 
 tightly bound states for positive z~parity but their absence for
 negative z~parity states. Transitions between
higher excited states which do not involve tightly bound
states, look different than their corresponding linear polarized
transitions: the wavelengths 
 increase or decrease as a function of $\gamma$. In the case of
 circular polarized 
transitions, states belonging to different magnetic quantum numbers $M$ are
involved and therefore the 
energy of the state with the higher absolute value of the magnetic
quantum number $M$  raises more strongly with increasing field
strength than the energy of the state with the lower magnetic quantum
number. This again goes back to the nuclear mass effects contained in
Eq.~(\ref{eq:trans}) and yields the increase or decrease of the wavelengths for
the circular polarized transitions with increasing field strengths.

In Fig.\ \ref{fig:trans0+to0-} the transition wavelengths for the singlet
and triplet transitions among the $\nu^{2S+1} 0^+$ and $\mu^{2S+1}
0^-$ states are
shown. The separation mentioned above, which is a general feature for 
linearly polarized transitions is obvious. The short
wavelengths are the transitions involving the 
most tightly bound state $1^10^+$. According to
Fig.\ \ref{fig:he_bind_m0}~(a) its ionization energy increases with
increasing field strength,
whereas the ionization energy of all other (excited) states varies
only marginally.

The circular polarized transition wavelengths shown in Fig.\
\ref{fig:trans0+to1+} involve the
$\nu^{2S+1} 0^+$ and $\mu^{2S+1} (-1)^+$ symmetry
subspaces. Separated by a large energetically gap there is a bundle of short 
wavelengths, which is due to transitions to the magnetically tightly
bound states $1^10^+$,$1^1(-1)^+$ and $1^3(-1)^+$  and a long
wavelength part.

The spectrum of circular polarized transitions among the subspaces $\nu^{2S+1}
0^-$ and $\mu^{2S+1} (-1)^-$, shown in Fig. \ref{fig:trans0-to1-}
shows also the characteristics of circular polarized transitions
discussed above. Since in this
case only states with negative z~parity are involved, we have no
tightly bound states,and the  part of the
spectrum with very short wavelengths is missing.

The behavior of the wavelengths of the linear polarized transition
from the $\nu^{2S+1}
(-1)^+$ states to the $\mu^{2S+1} (-1)^-$ states is shown in
Fig. \ref{fig:trans1+to1-} and
looks similar to 
the one given in Fig.\ \ref{fig:trans0+to0-}. 
Two well separated parts of the spectrum can be distinguished: Short
wavelengths, which are due to transitions to the magnetically tightly bound
states $1^1(-1)^+$ and $1^3(-1)^+$ and decrease
as a function of the  field strength.  Transition wavelengths
between higher excited states, with wavelengths larger than
1000~{\AA}ngstr{\o}m, remain approximately constant. Due to the fact, that
the influence of the finite nuclear mass is more significant than for those shown in
Fig.\ \ref{fig:trans0+to0-}, much less
lines belonging to bound-bound transitions are present.

The spectrum shown in Fig. \ref{fig:trans1+to2+} for the symmetry
subspaces $\nu^{2S+1}(-1)^+$ and $\mu{}^{2S+1} (-2)^+$ differs in some
respect from the
common pattern of 
circularly polarized transitions. Similar to other spectra  circular polarized
transitions, wavelengths involving transitions to the tightly bound states
($1^1(-1)^+$,$1^3(-1)^+$,$1^1(-2)^+$,$1^3(-2)^+$)
can be easily identified.  As a function of the magnetic
field strength, these wavelengths follow approximatively a power law. On the other
hand there are structures shown in Fig.\ \ref{fig:trans1+to2+}, which
arise  due to the influence of the finite nuclear mass. However in the gap
between these two parts of the spectrum,  additional
lines occur which are caused by  avoided crossings.
The transitions $\nu^{2S+1}(-1)^-$ to $\mu^{2S+1} (-2)^-$ shown in
Fig. \ref{fig:trans1-to2-} show the clear signature described above for
circular polarized transitions. Only a few lines belong to bound-bound
transitions. 

The spectral transitions given in Fig. \ref{fig:trans2+to2-} ($\nu^{2S+1} (-2)^+$
to $\mu^{2S+1} (-2)^-$) show the typical behavior
of linear polarized transitions. Deviating from the general pattern,
transition wavelengths to the state $2^1(-2)^+$
form their own bundle of short wavelengths, being located between the
wavelengths of  the transitions of
the tightly bound states $1^1(-2)^+$ and $1^3(-2)^+$ and the transitions
among higher excited states.

The spectrum of transitions among the subspaces  $\nu^{2S+1} (-2)^+$
to $\mu^{2S+1} (-3)^+$  is shown in Fig.\ \ref{fig:trans2+to3+}. 
Transitions to the magnetically tightly bound states
$1^1(-2)^+$,$1^3(-2)^+$, $1^1(-3)^+$, $1^3(-3)^+$ can be clearly
identified. Also the  transitions dominated by the  normal
finite mass effects  can be seen for long wavelengths. Additionally
the influence of the avoided crossings is visible.

\section{Brief Summary}
\label{sec:summ}

We have presented the first systematic full CI calculations for helium
in superstrong magnetic fields, taking into account the effects of finite
nuclear mass. These effects are extremely important in the superstrong
field regime, because the relevant parameter for the importance of the
finite nuclear mass effects
is $\gamma/M_0$. We
analyzed the influence of the normal and the specific finite nuclear
mass effects. It has been shown that the leading finite nuclear mass
effect, does not depend on the state, but only on
the magnetic quantum number. The state dependent part of the normal finite
 mass effects 
depends on the derivative with respect to the magnetic field strength
of the total energy of the corresponding state in
the infinite nuclear mass frame.  
Furthermore it has been shown, that the specific  mass
effects, which are caused by the mass polarization operators, are very
small compared to the total energies and small compared to the
leading normal  mass effects $\Delta E^e_{rm}$.

 In the superstrong magnetic field
regime, the spectrum of helium is cut off by the effects of the
finite nuclear mass. We found  that only a comparatively small number of
 
states is bound in the complete regime of  magnetic field strengths
investigated in the present work. Although the exact ionization threshold for
helium is unknown, all available approximations to the exact threshold confirm
this trend. Transition wavelengths for many linear and circular
polarized transitions were provided. Their typical behavior has been
identified and  the effects of the finite mass  on the transition
wavelengths has been analyzed.

The determination of the critical field
strengths (i.e. the field strengths where the individual
states become unbound) require a detailed investigation of the ground state
of the moving helium positive ion in a magnetic field.

\section*{Acknowledgments}
The Deutsche Forschungsgemeinschaft (O.A.A.) is gratefully
acknowledged for financial support.




\begin{thebibliography}{10}

\bibitem{Angel:1985_1}
J. Angel, J. Liebert, and H.~S. Stockmann, Astrophys. J. {\bf 292},  260
  (1985).

\bibitem{Angel:1978_1}
J. Angel, Ann. Rev. Astron. Astrophys. {\bf 16},  487  (1978).

\bibitem{Greenstein:1985_1}
J.~L. Greenstein, R. Henry, and R.~F. O`Connel, Astrophys. J. {\bf 289},  L25
  (1985).

\bibitem{Wunner:1985_1}
G. Wunner, W. R\"osner, H. Herold, and H. Ruder, Astron. Astrophys. {\bf 149},
  102  (1985).

\bibitem{Wickramasinghe:1988_1}
D.~T. Wickramasinghe and L. Ferrario, Astrophys. J. {\bf 327},  222  (1988).

\bibitem{Ruder:1994_1}
H. Ruder, G. Wunner, H. Herold, and F. Geyer, {\em Atoms in strong magnetic
  fields} (Springer Verlag, Berlin, 1994).

\bibitem{Kravchenko:1996_1}
Y.~P. Kravchenko, M.~A. Liberman, and B. Johansson, Phys. Rev. A {\bf 54},
  287  (1996).

\bibitem{Jordan:1998_1}
S. Jordan, P. Schmelcher, W. Becken, and W. Schweizer, Astron. Astrophys. Lett.
  {\bf 336},  L33  (1998).

\bibitem{Jordan:2001_1}
S. Jordan, P. Schmelcher, and W. Becken, Astron. Astrophys. {\bf 376},  614
  (2001).

\bibitem{Wickramasinghe:2000_1}
D.~T. Wickramasinghe and L. Ferrario, Pub. Astron. Soc. Pac. {\bf 112},  873
  (2000).

\bibitem{Mueller:1975_1}
R.~O. Mueller, A. Rau, and L. Spruch, Phys. Rev. A {\bf 11},  789  (1975).

\bibitem{Virtamo:1976_1}
J. Virtamo, J. Phys. B {\bf 9},  751  (1976).

\bibitem{Proeschl:1982_1}
P. Pr\"oschl, W. R\"osner, G. Wunner, and H. Herold, J. Phys. B {\bf 15},  1959
   (1982).

\bibitem{Vincke:1989_1}
M. Vincke and D. Baye, J. Phys. B {\bf 22},  2089  (1989).

\bibitem{Thurner:1993_1}
G. Thurner {\it et~al.}, J. Phys. B {\bf 26},  4719  (1993).

\bibitem{Becken:1999_1}
W. Becken, P. Schmelcher, and F. Diakonos, J. Phys. B {\bf 32},  1557  (1999).

\bibitem{Becken:2000_1}
W. Becken and P. Schmelcher, J. Phys. B {\bf 33},  545  (2000).

\bibitem{Becken:2001_1}
W. Becken and P. Schmelcher, Phys. Rev. A {\bf 63}, 053412 (2001).
\bibitem{Becken:2002_1}
W. Becken and P. Schmelcher, acc. f. publ. in Phys. Rev. A, (2001).

\bibitem{Pavlov:1995_2}
G.~G. Pavlov and A.~Y. Potekhin, Astroph. J. {\bf 450},  883  (1995).

\bibitem{Pavlov:1995_1}
G.~G. Pavlov, Y.~A. Shibanov, V.~E. Zavlin, and R.~D. Meyer,  in {\em Proc.
  {NATO} {ASI} {C} 450}, edited by M.~A. Alpar, U. Kizilo\v{g}lu, and J. van
  Paradijs (Kluwer, Dordrecht, 1995), pp.\ 71--90.

\bibitem{Taylor:1993_1}
J.~H. Taylor, R.~N. Manchester, and A.~G. Lyne, Astrophys. J. Suppl. Ser. {\bf
  88},  529  (1993).

\bibitem{Schmelcher:1994_1}
P. Schmelcher, L.~S. Cederbaum, and U. Kappes,  in {\em Conceptual Trends in
  Quantum Chemistry}, edited by E.~S. Kryachko and J.~L. Calais (Kluwer,
  Dordrecht, 1994), pp.\ 1--51.

\bibitem{Johnson:1983_1}
B.~R. Johnson, J.~O. Hirschfelder, and K.~H. Yang, Rev. Mod. Phys. {\bf 55},
  109  (1983).

\bibitem{Baye:1982_1}
D. Baye, J. Phys. B {\bf 15},  L795  (1982).

\bibitem{Baye:1986_1}
D. Baye and M. Vincke, J. Phys. B {\bf 19},  4051  (1986).

\bibitem{Vincke:1988_1}
M. Vincke and D. Baye, J. Phys. B {\bf 21},  1407  (1988).

\bibitem{Vincke:1990_1}
M. Vincke, J. Phys. B {\bf 23},  1991  (1990).

\bibitem{Baye:1990_1}
D. Baye and M. Vincke, J. Phys. B {\bf 23},  2467  (1990).

\bibitem{Schmelcher:1991_1}
P. Schmelcher and L. Cederbaum, Phys. Rev. A {\bf 43},  287  (1991).

\bibitem{Schmelcher:1995_1}
P. Schmelcher, Phys. Rev. A {\bf 52},  130  (1995).

\bibitem{Bezchastnov:1998_1}
V.~G. Bezchastnov, G.~G. Pavlov, and J. Ventura, Phys. Rev. A {\bf 58},  180
  (1998).

\bibitem{Chen:1992_1}
Z. Chen and S.~P. Goldman, Phys. Rev. A {\bf 45},  1722  (1992).

\bibitem{Poszwas:2001_1}
A. Poszwa and A. Rutkowski, Phys. Rev. A {\bf 63},  043418  (2001).

\bibitem{Avron:1978_1}
J.~E. Avron, I.~W. Herbst, and B. Simon, Ann. Phys. {\bf 114},  431  (1978).

\bibitem{Pavlov-Verevkin:1980_1}
V. Pavlov-Verevkin and B.~I. Zhilinskii, Phys. Lett. A {\bf 78 A},  244
  (1980).

\bibitem{Schmelcher:1988_1}
P. Schmelcher and L.~S. Cederbaum, Phys. Rev. A {\bf 37},  672  (1988).

\bibitem{Kappes:1996_1}
U. Kappes and P. Schmelcher, Phys. Rev. A {\bf 54},  1313  (1996); {\bf 53},
  3869 (1996); {\bf 51}, 4542 (1995).

\bibitem{Detmer:1997_1}
T. Detmer, P. Schmelcher, F. K. Diakonos, and L. Cederbaum, Phys. Rev. A {\bf 56},  1825  (1997).

\bibitem{Detmer:1998_1}
T. Detmer, P. Schmelcher, and L. Cederbaum, Phys. Rev. A  {\bf{57}}, 
1767 (1998); {\bf{61}}, 043411 (2000); {\bf{64}}, 023410 
  (2001); J. Chem. Phys., {\bf{109}} (1998); J. Phys. B {\bf{28}} (1995).

\bibitem{Al-Hujaj:2000_1}
O.-A. Al-Hujaj and P. Schmelcher, Phys. Rev. A {\bf 61},  063413  (2000).

\bibitem{Becken:2000_2}
W. Becken and P. Schmelcher, J. Comput. Appl. Math. {\bf 126},  449  (2000).

\bibitem{Loudon:1959_1}
R. Loudon, Am. J. Phys. {\bf 27},  649  (1959).

\bibitem{Ivanov:1998_1}
M.~V. Ivanov and P. Schmelcher, Phys. Rev. A {\bf 57},  3793  (1998); Phys.
  Rev. A {\bf{60}}, 3558 (1999); J. Phys. B {\bf{34}}, 2031 (2001); Eur. Phys.
  J. D {\bf{14}}, 279 (2001).

\end{thebibliography}
\end{document}